\documentclass[%
 rsi,
 amsmath,amssymb,
 reprint,%
]{revtex4-1}

\usepackage{graphicx}
\usepackage{dcolumn}
\usepackage{bm}

\usepackage[utf8]{inputenc}
\usepackage[T1]{fontenc}
\usepackage{mathptmx}
\usepackage{amsmath,amssymb} 
\usepackage{etoolbox}
\usepackage{booktabs} 
\usepackage[dvipsnames]{xcolor} 
\usepackage{acro} 
\usepackage{siunitx} 
\usepackage{enumitem} 

\makeatletter
\def\@email#1#2{%
 \endgroup
 \patchcmd{\titleblock@produce}
  {\frontmatter@RRAPformat}
  {\frontmatter@RRAPformat{\produce@RRAP{*#1\href{mailto:#2}{#2}}}\frontmatter@RRAPformat}
  {}{}
}%
\makeatother

 \DeclareAcronym{aLIGO}{short=aLIGO,long=Advanced Laser Interferometer Gravitational-wave Observatory}
 \DeclareAcronym{LIGO}{short=LIGO,long=Laser Interferometer Gravitational-wave Observatory}
 \DeclareAcronym{TM}{short=TM,long=test mass,long-plural=es}
 \DeclareAcronym{EM}{short=EM,long=electromagnetic}
 \DeclareAcronym{ESD}{short=ESD,long=electrostatic drive}
 \DeclareAcronym{EVMD}{short=EVMD,long=electrostatic violin mode damper, long-plural=s}
 \DeclareAcronym{DARM}{short=DARM,long=differential arm}
 \DeclareAcronym{QPD}{short=QPD,long=quadrant photodiode}
 \DeclareAcronym{SPD}{short=SPD,long=split photodiode, long-plural=s}
 \DeclareAcronym{PUM}{short=PUM,long=penultimate mass,long-plural=es}
 \DeclareAcronym{GW}{short=GW,long=gravitational wave,long-plural=s}
\begin{document}

\preprint{AIP/123-QED}

\title[Design of an electrostatic violin mode damper for gravitational wave detector enhancement]{Design of an electrostatic violin mode damper for gravitational wave detector enhancement}
\author{J. R. Rice}
\author{O. J. Weisenberger}
\author{A. R. Nadlman}
\author{G. A. R. Eddolls}
\author{T. C. Gayer}
\author{J. C. Lee}
\author{S. W. Ballmer}
\author{G. L. Mansell*}%
 \email{jrrice@syr.edu.}
\affiliation{ 
Center for Gravitational Wave Astronomy \& Astrophysics, Syracuse University
}%

\date{\today}

\begin{abstract}
The test mass suspension fiber resonances and their harmonics pollute the sensitivity of the detectors of the advanced Laser Interferometer Gravitational Wave Observatory (aLIGO) and cause operational issues.
The current method of damping these violin modes requires the GW detector be locked; identifying modes after a test mass replacement takes valuable commissioning time, and modes which are close together in frequency are hard to distinguish and damp independently. 
We present the design, prototyping, and initial lab testing of an electrostatic violin mode damper: a compact sensor and actuator set next to the fiber to monitor and damp resonances.
The design uses a split photodiode and LED integrated into a shadow sensor, and an electrostatic actuator. 
The device is vacuum compatible and its mechanical design is compatible with the suspension and surrounding cage of the advanced Laser Interferometer Gravitational Wave Observatory test masses.
The shadow sensor has a shot noise limited displacement noise floor of \SI{5e-11}{\meter\per\sqrt\hertz}, sufficient to measure the test mass violin modes when highly excited.
The modeled actuator strength is \SI{0.3}{nN}, meeting the requirements to damp the modes for the aLIGO suspensions.\\
\copyright 2026 Rice et al. This preprint is distributed under a Creative Commons Attribution-NonCommercial 4.0 International (CC BY-NC 4.0) license.
\end{abstract}

\maketitle

\section{\label{sec:intro}Introduction}

The \ac{aLIGO} detectors are a pair of dual-recycled Fabry-Perot Michelson interferometers which measure \ac{GW}-induced displacement of suspended \acp{TM} \cite{aLIGO}. 
The four \acp{TM} of each detector are \SI{40}{kg} fused silica optics, suspended from a quadruple suspension to isolate the \acp{TM} from ground motion.
The final stage of this suspension is monolithic fused silica: the \ac{TM} is suspended from the \ac{PUM} via four ultra-low loss fused silica fibers, minimizing the suspension thermal noise \cite{quad}. 
External excitations can ring up modes of the \ac{aLIGO} suspensions and, due to the fibers' extremely low mechanical loss, these modes can remain excited for many days \cite{vmode_glitchmon,vmode_kenstrain}.

Violin modes are transverse resonances of any string under tension with approximately fixed end points. 
The frequency of the violin modes can be estimated from the string/fiber's parameters and the tension it is under:
\begin{equation}
    f_n = \frac{n}{2L}\sqrt{\frac{T}{\mu}},
\end{equation}
where $f_n$ is the frequency of the nth harmonic, $L$ is the length of the fiber in meters, $T$ is the tension in the fiber in Newtons, and $\mu$ is the linear mass density of the fiber in kilograms per meter.
In reality the exact frequencies of the violin modes are affected by small deviations in the suspension fibers and their attachment to the test masses.
The \ac{aLIGO} suspension fibers have fundamental violin mode frequencies at $\sim$\SI{500}{Hz} and harmonics \cite{lockerbie_first_2011}.
These modes are in the \ac{GW} detection band, however much of the energy of the mode is focused into a spectrally narrow peak. 
Violin modes couple into detector displacement noise by producing a small recoil in the \ac{TM} due to their motion \cite{cumming}. 
These sharp resonant peaks reduce the sensitivity of the detector by many orders of magnitude within the detection band, with the exact violin mode amplitude depending on external excitation \cite{O4_lines}.
The noise from the violin modes covers potentially interesting astrophysical signals including parts of the signal waveforms from compact binary coalescences, and potential continuous wave sources \cite{Abac_2026_cw}.
Currently the modes are sensed with the gravitational-wave detector readout, and damped by feeding back to the penultimate mass.

Damping of the violin modes can be difficult.
Each suspension fiber has two fundamental resonant frequencies.
The splitting of these two violin modes appears in part to originate from asymmetries in the neck between stock and fiber, see Appendix \ref{app:frequencies}.
For modes that vibrate in the direction of the interferometer arm, it is possible to sense the mode due to the direct coupling into the displacement of the \ac{TM}.
The orientation of the mode axes varies from fiber to fiber.
With four fibers per \ac{TM} and four \acp{TM} per detector, resolving the modes of each \ac{TM}, which can be extremely close in frequency \cite{vmode_data_comp}, can be time-consuming. 
The current violin mode sensing method requires the interferometer be locked and operational, as the sensor is the \ac{GW} readout, resulting in valuable commissioning time devoted to damping modes and identifying modes and damping settings after a test mass replacement.
Furthermore, damping of modes perpendicular to the arm transverse axis can be extremely difficult or impossible due to the weak coupling to the GW readout, however through complex cross-couplings such modes can still impinge on the detector's sensitivity.

The goal of this work is to produce a device capable of individually sensing and actuating on the fused silica fibers.
This will allow the quick identification of violin modes of each fiber, reducing commissioning time spent attributing each mode to a test mass and providing new insight into the violin mode splitting.
The ability to sense the modes without relying on a locked interferometer will enable faster damping and reduce downtime after external excitation events such as earthquakes.
We present a compact shadow sensor combined with an electrostatic actuator, similar in principle to the \ac{aLIGO} \acp{ESD} \cite{ESD}, to sense and actively damp the violin modes of each suspension fiber independently.
The design presented here targets the \ac{aLIGO} \acp{TM}, but will be easily adapted to new heavier suspension upgrades to \ac{aLIGO} and next generation detectors \cite{postO5,ce_mpsac}.

The structure of the paper is as follows:
Section~\ref{sec:design} describes the engineering constraints, design requirements, 
and our resulting device design of the \ac{EVMD} including the mechanical, sensor and actuator designs.
Section~\ref{sec:results} shows results of the initial testing of the \ac{EVMD} in a tabletop experiment including a shorter, smaller radius fused silica fiber.
Section~\ref{sec:conclusion} summarizes the current status of the violin damper and explores future work on the path to install in \ac{aLIGO}, as well as adaption of this design to future \ac{GW} detectors and upgrades.
\section{Design}
\label{sec:design}
This section summarizes the design of the electrostatic violin mode damper (EVMD).
Section~\ref{subsec:req} describes the requirements of the device for integration into \ac{GW} detectors and for sensing and actuating on violin modes. 
Section~\ref{subsec:mech} describes the resulting mechanical design; Section~\ref{subsec:sensing} describes the design of the violin mode sensor; Section~\ref{subsec:actuator} describes the actuator design.

\subsection{Requirements}
\label{subsec:req}
The \ac{EVMD} must be able to sense and damp the resonances of two \SI{400}{\micro\meter} diameter fused silica suspension fibers.
For aLIGO, the suspension fibers are spaced \SI{3}{\centi\meter} apart \cite{quad}, and are protected by a fiber guard with internal cross-section of {\SI{55}{\milli\meter} $\times$ \SI{41}{\milli\meter}} \cite{fiberguardmainbody,fiberguard, fiberguardinstall}, with an internal guard between the two fibers, constraining the available space for the \ac{EVMD}.
To sense and actuate on the modes of each fiber, two sets of sensors and dampers must match the fiber spacing and be resilient to reasonable tolerances around this value.
Each fiber has two fundamental resonances along orthogonal axes; the orientation of these axes is random with respect to the test mass.
An investigation into the violin mode directions is presented in Appendix \ref{app:frequencies}, preliminary evidence suggests the mode axis alignment is tied to asymmetries in the fiber neck.
The inner part of the \ac{EVMD} is able to rotate about the fiber, allowing the alignment of the shadow sensors and actuators to the two fundamental violin modes during installation.
Any cross couplings between sensors and actuators of neighboring fibers must be mitigated.
To allow for damping of higher order resonances, the vertical position of the \ac{EVMD} along the fiber is \SI{50}{mm} above the center of the fiber, making use of a window in the fiber guard intended for a previous violin mode sensor that was ultimately not installed in \ac{aLIGO} \cite{lockerbie_first_2011}.
This placement is designed to coincide with a null of the 5th order violin mode, though the precise node location depends on the individual fiber profile.
The \ac{EVMD} will be sensitive to all lower order modes which are of most interest to \ac{aLIGO}.

Any components to be integrated into the \ac{aLIGO} detectors must adhere to strict requirements.
The violin mode damper must not degrade the exquisite sensitivity of the \ac{GW} detector and any noise contributions to the \ac{GW} detector readout must be a factor of 10 below the design sensitivity.
The violin damper must also be compatible with ultrahigh vacuum, and any components or materials not included on the \ac{aLIGO} vacuum materials list \cite{ligo_vac_materialse} must be tested for outgassing in a residual gas analyzer and approved for use prior to integration.
Given the \ac{EVMD}'s necessary close proximity to the suspension fibers, the mechanical design and install procedure must maintain a \SI{3}{\milli\meter} buffer zone around the fiber to ensure no contact with the fiber.
Moving parts, gears, and threads are minimized to reduce the risk of debris dropping from the damper onto the test mass and/or fibers.
The design must be mechanically simple and easy to install on the existing suspension cage hardware.
Finally, the design must include as few electrical connections as possible, and stray electric fields from the sensor and actuator must be at least a factor of 10 below the \ac{GW} detector design sensitivity.

When the violin modes are highly excited, fiber displacement can reach more than \SI{7.5e-11}{m} RMS, setting the minimum sensitivity requirement of the \ac{EVMD} sensor.
This displacement can be estimated from the violin mode peak height measured in the \ac{GW} detector readout, and scaling by the mass ratio of the test mass to the fiber mass. 
A more rigorous calculation is shown in Appendix~\ref{subsec:sensingest}.
To damp the modes down to the \ac{GW} detector noise floor underneath the violin modes, a sensitivity of \SI{e-16}{m} RMS is required.
This second more stringent noise floor is extremely challenging to achieve in a compact sensor.
If the second noise floor cannot be achieved, the violin damper is envisioned to be used in conjunction with existing violin mode damping schemes to do the final low-noise damping of the modes.

\begin{figure}[h]
\includegraphics[width=\columnwidth]{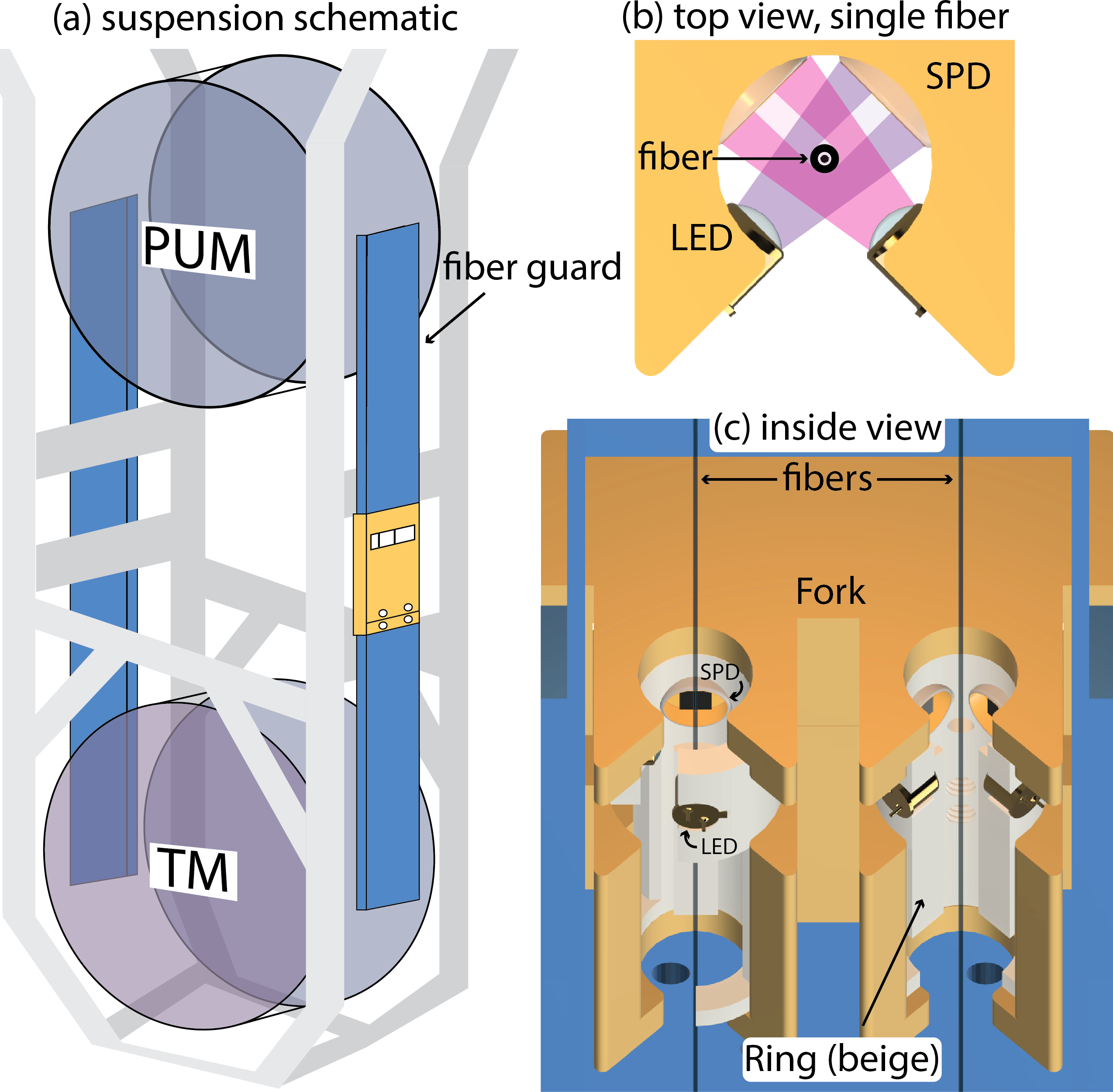}
\caption{Schematic of the \ac{aLIGO} suspension showing the location of the \ac{EVMD}, the alignment of the shadow sensors and fiber, and CAD model of the \ac{EVMD} components.
(a), left, shows a schematic of the two bottom stages of the \ac{aLIGO} suspension. The test mass (TM) is the bottom stage, suspended from the penultimate mass (PUM).
The general structure of the suspension stage is shown in gray, the fiber guard is shown in blue, and the \ac{EVMD} is in yellow.
(b), top right, shows a top-down view of the sensors around a single fiber. Orthogonal LEDs illuminate the fiber and the shadows are centered on two split photodiodes (SPDs).
The beams from the two LEDs shown in different colors for clarity, but are the same wavelength.
(c), bottom right, shows a CAD model rendering of the \ac{EVMD} positioned around the two neighboring fibers. 
The sensors around the right fiber are in the 'x' alignment, as they will be installed, while the sensors around the left fiber are in the '+' alignment.
The sensor orientation with respect to individual fibers is set to maximize each sensors sensitivity to a single violin mode resonance. 
The alignment will be finalized during the installation of the \ac{EVMD}.}
\label{fig:evmd}
\end{figure}

To damp the modes, an actuation force of \SI{3.2e-13}{N} RMS is required at the fiber antinode. 
This requirement is set by the energy stored in the mode, the target energy damping time of \SI{60}{s}, and the work done per cycle by the actuator; 
the estimate is derived in Appendix~\ref{subsec:dampingest}.

\subsection{Mechanical design}
\label{subsec:mech}

To address the mechanical requirements discussed above, the \ac{EVMD} is designed to slot into an existing window in the fiber guard, as shown in the schematics in Fig. \ref{fig:evmd}. 
Figure \ref{fig:evmd}a. shows a schematic of the bottom stages of the \ac{aLIGO} test mass suspension (left) with the \ac{EVMD} fork, which will replace an existing window cover, shown in yellow.
The \ac{EVMD} sensors and actuators are mounted in a PEEK ring which, after install, may be rotated around the fiber to optimally align the shadow sensor and actuator to the violin mode axes - detailed view in Fig. \ref{fig:evmd}b.
Each unit has two rings which are centered on neighboring fibers.
The rings are held in a fork which slots into the fiber guard window (the fork is shown in yellow in Fig. \ref{fig:evmd} on the right).
On the back of the fork, slotted holes allow for millimeter scale adjustments in the position of the fork relative to the fiber guard.
This is necessary as the spacing between the fiber guard and the fiber is set by eye, with millimeter tolerancing.

During install, the rings are held in a "x" alignment such that the \ac{EVMD} can be installed without impacting the fiber, as demonstrated in the rightmost fiber in Fig. \ref{fig:evmd}c.
The position of the fork is adjusted while monitoring the shadows on the split PDs in real time to ensure the initial alignment.
After the translational alignment, the rings can rotate inside the fork allowing the shadow sensors to be aligned with the fiber mode axes.
The modes can be deliberately excited using a speaker during this rotational alignment.
The \ac{EVMD} can be rotated $\pm 45^{\circ}$, the left fiber shown in Fig. \ref{fig:evmd} shows a sensor in the "+" alignment.
This allows the \ac{EVMD} to be configured so each mode of the fiber is predominantly sensed and actuated by a single sensor and actuator.
Once the alignment of the \ac{EVMD} is set, the rings are secured to prevent misalignment during pump down to vacuum.

\begin{figure}
\includegraphics[width=0.85\columnwidth]{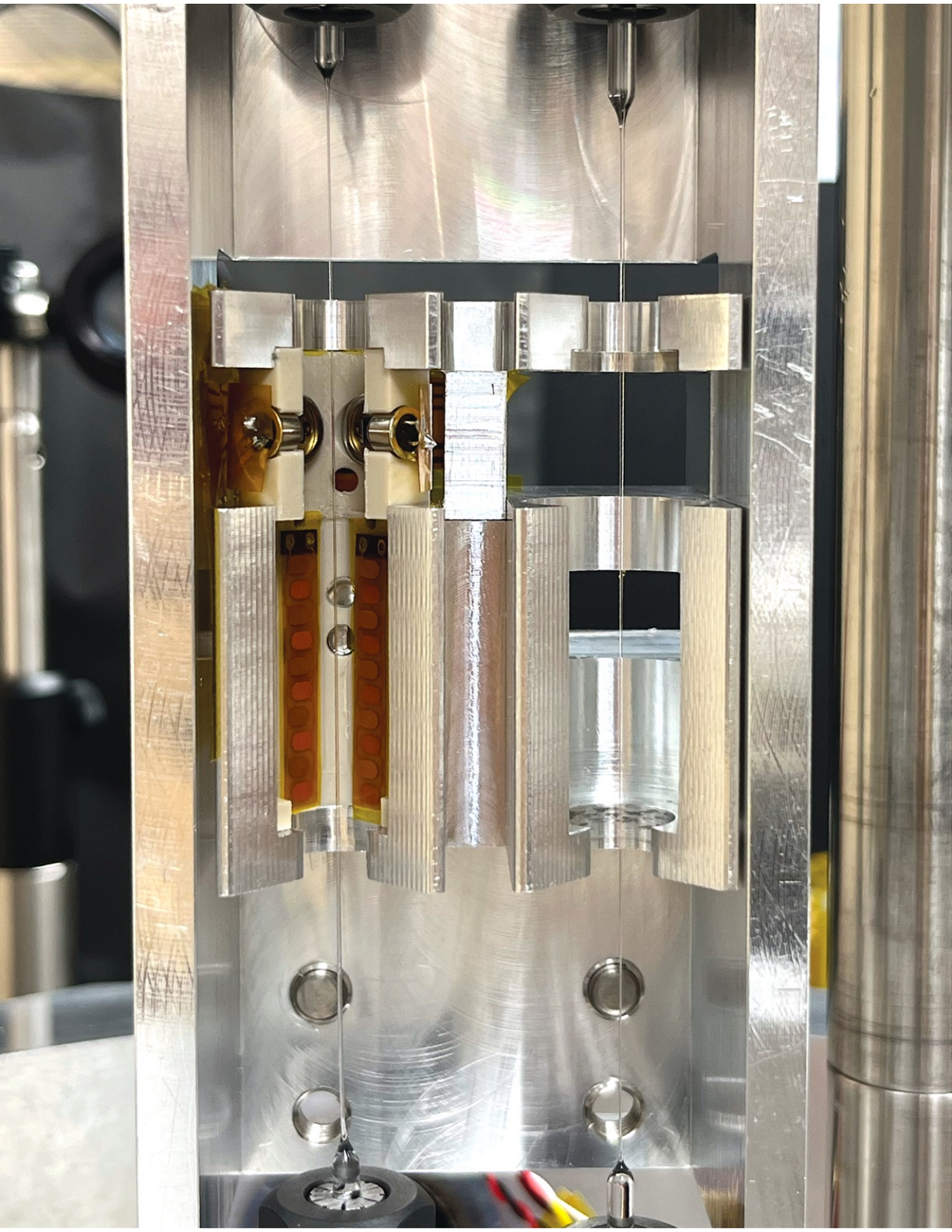}
\caption{Photo of the \ac{EVMD} fork and single ring in a test set up.
Two fibers, separated by \SI{3}{\centi\meter} suspend a mass.
The right fiber shows the fork without the ring and electronic components installed.
The left fiber has \ac{EVMD} shadow sensors in the x configuration (LEDs closest, SPDs partially obscured) and flexicircuit actuators.}
\label{fig:photo}
\end{figure}

A photograph of the initial \ac{EVMD} prototype is shown in Fig. \ref{fig:photo}. 
In this initial build only one ring is included.

\subsection{Sensor design}
\label{subsec:sensing}
To sense the violin modes, the \ac{EVMD} monitors the fiber displacement directly using two shadow sensors.
The design is comparable to existing shadow sensor devices used in \ac{GW} detectors \cite{carbone, BOSEM, akutsu}.
The design also draws on a previous prototype of a violin damper \cite{lockerbie_tia,lockerbie_violin-mode_2014}.

Two vacuum compatible LEDs (Hamamatsu L11913, $\SI{850}{nm})$ illuminate the fiber along orthogonal axes. 
The fibers cast a vertical linear shadow, which is centered on a two-element split photodiode (PD) (Advanced Photonix).
To maintain the necessary \SI{3}{\milli\meter} buffer region around the fiber and provide additional space for sensor alignment, the LED and PD are separated by \SI{10}{\milli\meter} along the LED's optical axis.
As the fiber oscillates at the violin mode frequencies, the shadow is displaced from the center of the split PD, inducing more photocurrent in one of the sensors, and less in the other.
The photocurrents of each PD segment are sent to independent transimpedance amplifiers, converting the photocurrents to voltages. 
A differential amplifier then subtracts the segment voltages yielding a signal at the violin mode frequencies.
The LED was chosen as it is already \ac{aLIGO}-vacuum approved, and used in similar shadow-sensor devices in the detectors.

\begin{figure*}[t] 
    \centering
    \includegraphics[width=\textwidth]{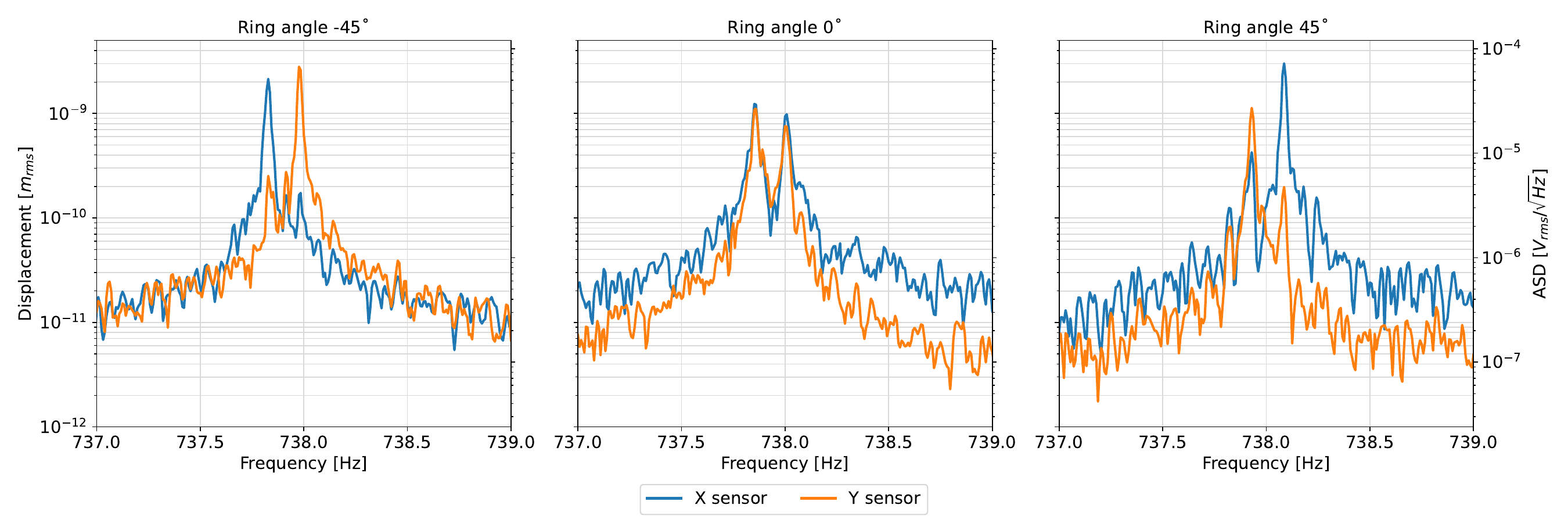} 
    \caption{Violin modes of the shorter and thinner test fiber in the \ac{EVMD} testbed measured with orthogonal sensors (blue and orange traces) at three different angles (left, middle and right subplots).
    At $\pm 45^{\circ}$, the x- and y- sensors are closely aligned with the fiber's violin modes; at $0^\circ$, both sensors have partial sensitivity to both modes.
    The right y-axis shows the amplitude spectral density of the \ac{SPD} difference voltage; 
    left y-axes are calibrated into root-mean-square (RMS) fiber displacement, obtained by multiplying the sensor spectrum by the square root of the effective noise bandwidth (ENBW) \cite{DFT} and dividing by the sensor calibration of 2120 V/m (Fig. \ref{fig:calib}).
    These measurements were performed with a Stanford Research Systems SR785 with 4, 3, and 2 averages respectively.
    A \SI{6.25}{Hz} span, 800 FFT lines (\SI{7.8}{mHz} bin spacing) and a Blackman-Harris window were used, giving an ENBW of \SI{15.6}{mHz}. 
    This data was taken at rough vacuum to mitigate air damping, however exact vacuum levels varied from measurement to measurement resulting in slight shifts in the violin mode frequencies.
    Vibrational coupling and number of averages varied between measurements affecting the sensor noise floors.}
    \label{fig:full_width}
\end{figure*}

The performance of the prototype sensors in the test set up is shown in Fig. \ref{fig:full_width}, which shows the sensitivity to violin mode oscillations dependence on angle of the shadow sensors.

\begin{figure} [h!]
\includegraphics[width=\columnwidth]{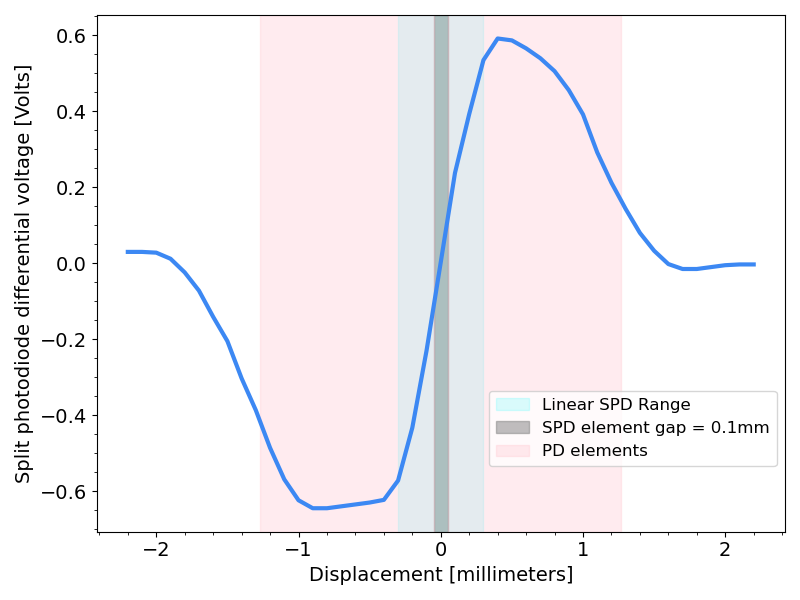}
\caption{Calibration of the EVMD sensor in the test set up: split photodiode (SPD) differential voltage $\left(V_{segment~1} - V_{segment~2}\right)$ versus \ac{SPD} displacement. 
The scale of the photodiode elements are shown in pink shading, the gap between PD elements is shown in gray, the linear response region is shown in teal.
The blue curve shows the response of the SPD to shadow displacement and serves as both a measure of the photodiode linear range and a calibration from voltage to displacement.
The slope of the linear region yields a calibration of \SI{2120}{\volt/\meter}.
Note the fiber used here is \SI{250}{\micro\meter} in diameter, a factor of two below the suspension fibers used in \ac{aLIGO}. 
The larger fiber will cast a larger shadow, increasing the region of linear response of the sensor to fiber displacement.}
\label{fig:calib}
\end{figure}

To ensure the EVMD has sufficient sensitivity to measure the violin modes in \ac{aLIGO}, the prototype is calibrated into displacement sensitivity.
To calibrate the sensor, the shadow is translated across the split PD by translating the PD with a micrometer.
The DC voltage from the split PD as a function of shadow translation gives the calibration from Volts to Meters.
Figure \ref{fig:calib} shows the calibration curve; for maximum sensitivity, the LED-fiber-PD axis must be aligned such that the shadow is centered on the gap between PD elements. 
This alignment tolerance is set by the linear response region in Fig. \ref{fig:calib}.

Measurements of the sensor's typical noise floor are approximately \SI{1e-7}{\volt_{rms}\per\sqrt\hertz} in the audio band, which is consistent with the expected shot noise given the DC signal from the individual photodiode segments.  
This calibrates into a displacement noise floor of \SI{5e-11}{\meter\per\sqrt\hertz}.
Thus fiber motion exceeding this level at the violin mode frequencies is detectable with this sensor. 
This measurement was made using a smaller fiber diameter, with full experimental description in Section \ref{sec:results}. 
A similar sensitivity is expected with the \ac{aLIGO} fibers, and their larger shadow is anticipated to increase the dynamic range of the sensor. 

\subsection{Actuator}
\label{subsec:actuator}

Once the modes are sensed with the shadow sensor, the signal is digitized and filtered in the \ac{aLIGO} digital control system (CDS) \cite{BORK2021100619}, in the same way the sensed modes are processed in the current damping method.
A phase-shifted control signal will be sent to the \ac{EVMD}'s electrostatic actuator in order to damp the modes.
The actuator design for the \ac{EVMD} draws on the electrostatic drive (ESD) used in \ac{aLIGO} \cite{ESD}.
Two electrodes -- the DC bias and the signal at the violin mode frequency -- are arranged in an interleaving comb. 
The bias electrode is held at a DC  voltage to induce charge polarization in the fused silica fiber.
The signal at the violin mode frequency is sent to the signal electrode.

To actuate on the fiber with sufficient force as to damp the modes, \SI{3.2e-13}{N} of actuation force is required. 
The force (F) from an electrostatic actuator is given by 

\begin{equation}
F = \alpha (\Delta V)^2
\end{equation}

Where $\Delta V = | V_{s} - V_{b} |$ \cite{ESD2}. Here, $V_{s}$ is the signal voltage, $V_{b}$ is the bias voltage, and $\alpha$ is a proportionality factor dependent on the actuator parameters including length, periodicity and width of the electrodes, and distance to the target of actuation.
A rigorous definition and derivation is shown in Appendix B. 

\begin{figure}
\includegraphics[width=\columnwidth]{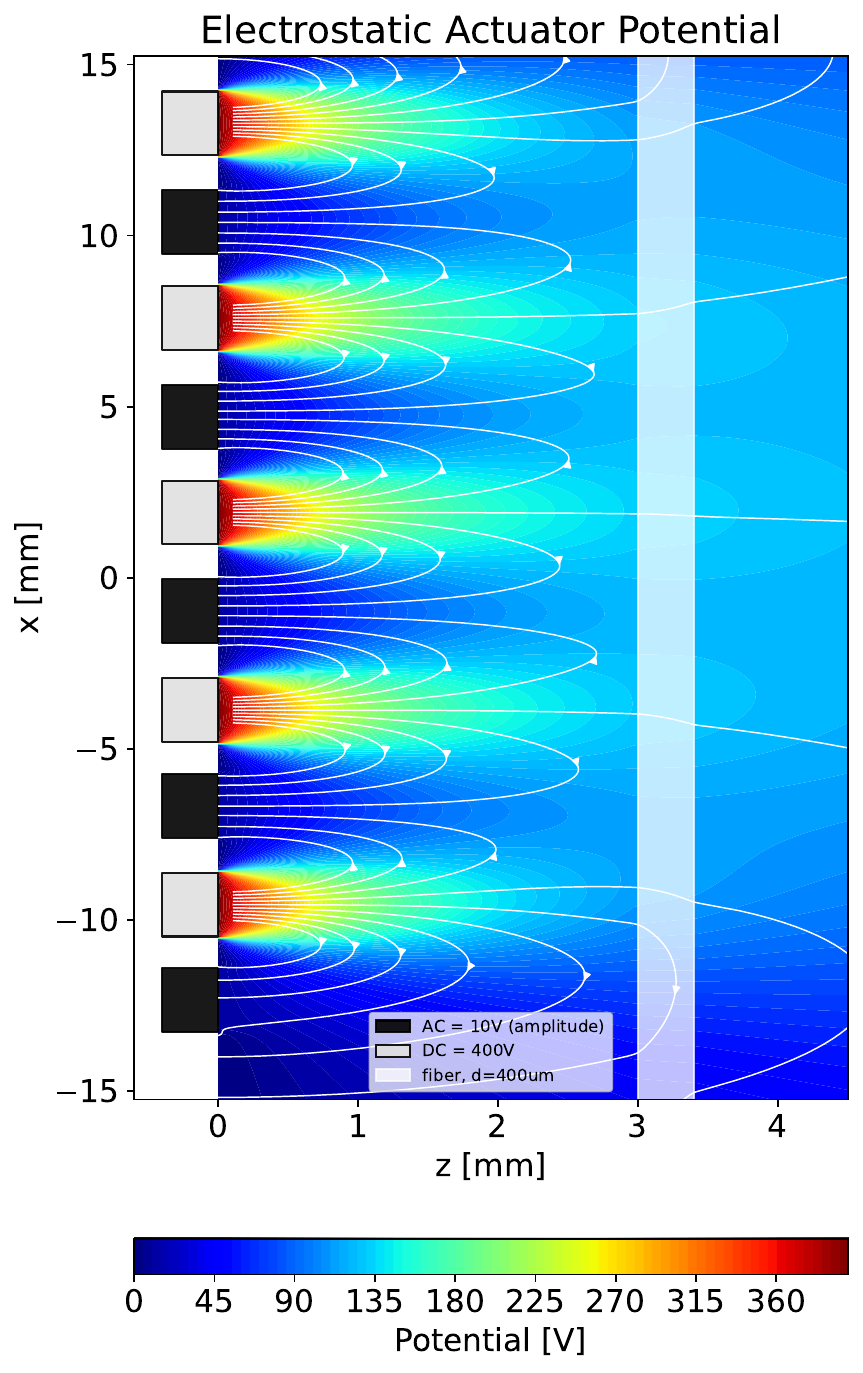}
\caption{Modelled electrical potential around the \ac{EVMD} actuator. 
Left side shows actuator periodicity, with signal electrode in black and bias electrode in gray.
Color gradient shows the electrical potential with fringing field lines shown in white. 
Fiber is shown at its nominal distance from the actuator.}
\label{fig:actuator}
\end{figure}

To determine the optimal geometry of the actuator, including electrode width and spacing, analytical and ANSYS modeling was used to analyze electric fields around the actuators.
One of the tested actuator geometries and the analytically modeled electric field is shown in Fig. \ref{fig:actuator}.
The models verify the actuation strength of each configuration, while the 3D ANSYS model additionally checks for stray out-of-plane fields and other technical issues with actuator design.

To maintain acceptable distance between the fiber and \ac{EVMD}, the actuator must be on the order of \SI{0.1}{\milli\meter}.
As shown in Fig. \ref{fig:actuator}, the thin actuator is achieved by encapsulating the electrodes in a Kapton casing (Flexicircuit). 
Flexicircuit is vacuum compatible, and an additional Flexicircuit wrapped around the outside of the rotation ring is used to connect the \ac{EVMD} electronic components (LED, split PD, and actuator) to a connector on the back of the device.
Prior to testing under \ac{aLIGO} conditions, these Flexicircuits will be replaced with alternative Kapton-encased circuitry that is compatible with \ac{aLIGO}'s ultrahigh vacuum standards.

\section{Results and discussion}
\label{sec:results}

A prototype of the \ac{EVMD} has been constructed and tested in a tabletop experiment, see Fig. \ref{fig:photo} for a photo of the experiment.
This section describes the build and integration of the \ac{EVMD} assembly and test set up.
Finally, we present measurements of the first harmonic violin modes, measured as a function of angle around the fiber.

The \ac{EVMD} mechanical components were machined in-house at Syracuse University.
The LEDs and split PDs are press-fit in the vacuum-compatible PEEK ring, which is in turn held in an aluminum fork.
Initial alignment to ensure the beam from the LED is centered on the split PDs is completed before installation. 
With the LEDs on, small tweaks in their alignment are completed while monitoring the subtraction of the two elements of the split PDs. 
In future builds, this alignment will be finalized and the components secured with UV-cured epoxy; in this initial build Kapton tape was used.

The flexi-circuit actuator is mounted in the ring placed below the split photodiodes.
The electrical connections to these components are routed through a flexicircuit strip wrapped around the back of the PEEK ring.
A D-sub connector is mounted behind the ring on the fork.
The fork interfaces with the fiber guard and holds the two rings around the two neighboring fibers, see Fig. \ref{fig:evmd}.

The test set up consists of two fibers suspended from a breadboard table housed in a bell-jar vacuum tank.
These test fibers, pulled at the University of Glasgow, are \SI{110}{mm} in length and have diameters of \SI{250}{\micro\meter} which is approximately half the diameter of the fibers used in the \ac{aLIGO} suspensions.
Thinner fibers were chosen to keep the violin modes of the fibers in the test set up at the same order of magnitude of the violin modes in \ac{aLIGO}.
A short fiber guard, with the same cross-sectional dimensions and window feature as the \ac{aLIGO} fiber guard, is mounted to the breadboard table, and the \ac{EVMD} slots into the fiber guard.
To suspend the fibers, the glass-to-metal interfaces are made by clamping the fiber stock in machining collets.
The top collets are attached to the underside of the breadboard table, the bottom collets are clamped to a suspended \SI{400}{\gram} mass.

The \ac{EVMD} is assembled outside the tank; with the rings held in the "x" configuration, the \ac{EVMD} is slotted around the fiber.
Initial alignment of the \acp{EVMD} to the fibers is completed with the LEDs on, monitoring the subtraction of the two elements of the split photodiodes.
The shadow must be well-centered on the split photodiode to ensure maximum sensitivity to the fiber displacement.
To align the \ac{EVMD} transverse to the fibers, slotted holes in the back of the fork allow millimeter scale adjustments.
For alignment of the \ac{EVMD} into the fiber guard, shims were used in this initial build to set the offset between the \ac{EVMD} and fiber guard. 
This initial build made clear the need for independent alignment of each ring due to offsets in the fiber necks introducing deviations from the 3 cm separation between adjacent fibers. 
In the future version of the \ac{EVMD}, this will be achieved by building in tolerance to the spacing between the ring and the fork.

With the test set up, the initial alignment process of the LED, fiber and split PD is verified.
The sensor noise floor is characterized, actuation strength modelled, and the violin modes of the fiber are characterized.
Figure \ref{fig:full_width} shows measurements of the two violin modes of a single fiber measured at three different angles of the ring inside the fork.
The individual modes are orthogonal, and the sensor is aligned to them before pumping down to vacuum to take each measurement.

\section{Conclusions and Future work}
\label{sec:conclusion}

We have presented a compact shadow sensor and electrostatic actuator to measure and damp the violin modes of the \ac{aLIGO} suspensions.
The \ac{EVMD} is designed to be mechanically compatible with the existing \ac{aLIGO} suspensions, and is \ac{aLIGO} vacuum compatible.
We have successfully tested the installation procedure and alignment of an \ac{EVMD} prototype in a test setup including a mini fiber guard and fused silica fibers.

Each fiber has two violin modes which the sensor - consisting of two LEDs and split photodiodes - is aligned to after installation.
Measurements from the \ac{EVMD} prototype show that the sensor has a noise floor of \SI{5e-11}{\meter\per\sqrt\hertz} - low enough to measure the violin modes in \ac{aLIGO} when they are in an excited state. 
The \ac{EVMD} will thus allow for faster recovery after modes are rung up from external impulses including earthquakes.
To fully damp the modes to the \ac{GW} detector noise floor, the \ac{EVMD} must be used in conjunction with the \ac{GW} detector readout.

The electrostatic actuator is designed to have sufficient force at the fiber to damp the violin modes.
Analytic and FEA modelling of a variety of actuator geometries informed the current actuator design, which has a modeled actuation strength on the order of a few hundred picoNewtons.
Estimates of the voltage noise and stray electric field coupling to \ac{DARM} are future work.

The current version was designed anticipating an exact separation of \SI{30}{\milli\meter} between the fibers.
However, it is now apparent that the fiber separation can vary by several hundred microns or more depending on asymmetries in the fiber necks, see appendix A. 
Future versions of the \ac{EVMD} will add fine control of the separation of the two rings inside the fork, to allow alignment to the individual fibers.
Additional future additions to the \ac{EVMD} include actuators with improved spacing and length to increase actuation strength and the use of UV-cured epoxy to adhere LEDs and SPDs to the rings after a careful alignment.

Future measurements include successful sensing and damping of the modes of two adjacent fibers in the tabletop set up, after which the integration of the \ac{EVMD} will be demonstrated on a spare \ac{aLIGO} suspension.
Finally, the \ac{EVMD} will undergo modifications to be compatible with future \ac{aLIGO} upgrades \cite{postO5} and next-generation \ac{GW} detectors \cite{ce_mpsac}.
These suspensions are currently in a design phase, however thicker fibers with larger spacings are anticipated, parameters to which the \ac{EVMD} will easily be adapted.

\section{Acknowledgements}

We thank Alan Cumming and the Institute for Gravitational Research at the University of Glasgow for helpful discussions and for providing fibers for the test set up.
We thank Thomas Roocke and David Pratt for helpful discussions.
This work was supported by the U.S. National Science Foundation award PHY-2409519.


\appendix

\section{Violin mode splitting and directionality}
\label{app:frequencies}

\begin{figure}
\includegraphics[width=\columnwidth]{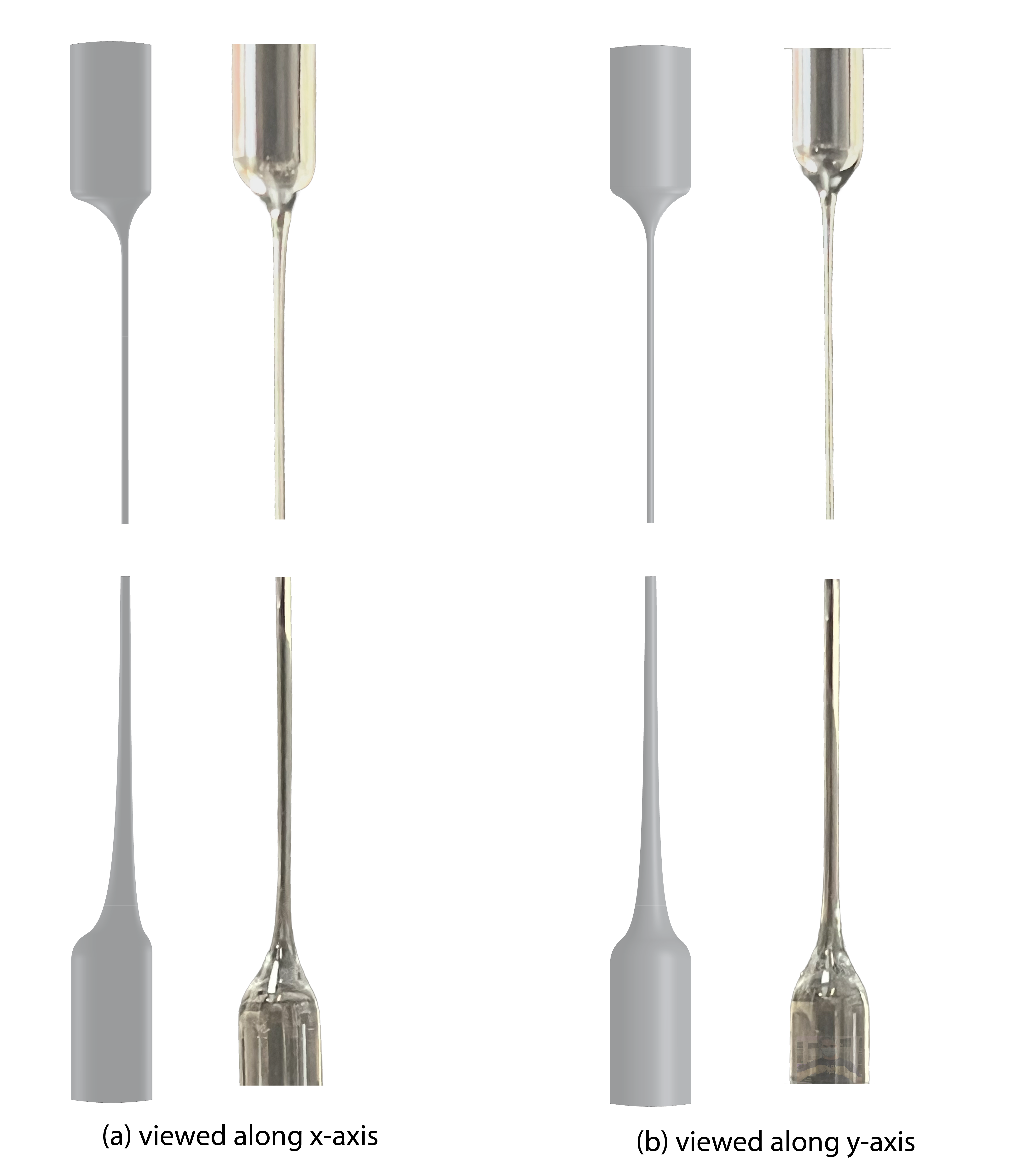}
\caption{Comparison of fiber neck asymmetry model and photograph.}
\label{fig:mode_axes}
\end{figure}

In both the EVMD experimental setup and the aLIGO suspensions, each fiber has orthogonal oscillation axes that manifest as two fundamental resonant frequencies. 
These two fundamental modes, sometimes referred to as ``mode splitting" could be partially attributed to several asymmetries that exist in the geometry of the fibers' tapered ``neck" regions.
The \ac{EVMD} experimental fibers exhibit a slight offset between the length of the thinnest part of the fiber and the central axis of the stalk. 
This offset, originating at the neck of the fiber, can be on the order of \SI{0.1}{\milli\meter}.
Additionally, the fibers taper more gradually on one end and more rapidly on the other, resulting in one side of the fiber being slightly thicker near the neck.
Two separate fibers have been used in the EVMD experimental setup, each displaying these asymmetries to different extents.
A relationship has been observed between the fibers' geometric discrepancies and the magnitude of their mode splitting:
The fiber with a visibly larger offset in the neck region (shown in Fig. \ref{fig:mode_axes}) has a mode separation of ~\SI{10}{\hertz}, as compared to a fiber with a much smaller offset, having a mode separation of ~\SI{200}{\milli\hertz} (see data in Fig. \ref{fig:full_width}).  
For the fiber with a \SI{10}{\hertz} mode separation, ellipticity in the length has been ruled out as the primary source of mode splitting through careful analysis of the fiber profiles and FEA modeling.

Ansys FEA modal analysis was used to investigate this relationship.
3D models of a fiber were developed using the same diameter and length of the fiber in the experimental setup.
Geometric asymmetries were added to the fiber models to approximately match those of the more asymmetric fiber in the experiment, including neck offsets and a thicker taper on one side. 
Figure \ref{fig:mode_axes} shows a comparison of the modeled fiber and real fiber.  
Both the real and modeled fibers have elliptical cross sections along the neck regions and a circular cross sections along their length.
Ansys modal analysis confirmed the relationship observed in the experimental fibers, with greater geometric asymmetry in the neck producing larger mode splitting. 
Results of this modeling are summarized in Table \ref{tab:fiber_modeling}.
Modal analysis was also performed in SolidWorks for further confirmation, showing good agreement with the Ansys results.
The modeling further suggests that the directionality of the modes also depends on this asymmetry, 
with one mode consistently oscillating along the axis of asymmetry and the other along the perpendicular axis.
Directly attributing the oscillation axes to the asymmetry of experimental fibers is more challenging, as the full complexity of their neck geometry is not captured by the 3D models. 
However, characterization of the experimental fibers still suggests their oscillation axes are related to asymmetry in the neck region;
the axes remain fixed with respect to the fiber itself, independent of how it is clamped or oriented in the suspension.
This remains true even when two fibers are suspended together, similar to the aLIGO quadruple suspensions where two fibers are used on either side of the optic. 

More testing is needed to determine if the \ac{aLIGO} fibers demonstrate a similar relationship between geometric asymmetries, mode splitting, and mode directionality.
Several important discrepancies also exist between the fibers used in this experiment and those employed on the GW detector. 
The fibers in this experiment are under significantly less stress (\SI{40}{\mega\pascal}) than those in \ac{aLIGO} (\SI{800}{\mega\pascal}) \cite{PhysRevApplied.17.024044_3g_fibers}.
The \ac{aLIGO} fibers also have additional geometric features (and thus potential for more asymmetries) that need to be considered: 
the fibers have a thermoelastic loss cancellation region between the fiber neck and the length of the thinnest part of the fiber \cite{Bell_2014_thermoelastic}, and they are also directly welded to the fused silica "ears" of the suspended optic.
FEA modeling of these asymmetries is an area of ongoing research. 

\begin{table}[htbp]
    \centering
    \begin{tabular}{l r}
        \toprule
        Modeled asymmetry & Mode separation (\si{\hertz}) \\
        \midrule
        None (perfectly symmetric fiber) & \num{0.04} \\
        \SI{0.2}{\milli\meter} neck offset & \text{0.16} \\
        \SI{0.5}{\milli\meter} neck offset & \text{3.12} \\
        \SI{0.5}{\milli\meter} neck offset and thicker bottom taper & \text{11.76} \\
        \bottomrule
    \end{tabular}
    \caption{Mode separation resulting from asymmetries added to a 3D-modeled fiber in Ansys modal analysis.}
    \label{tab:fiber_modeling}
\end{table}

\section{Estimations of sensing and damping requirements}
\label{sec:estimates}

\subsection{Sensing requirement estimate}
\label{subsec:sensingest}
The minimum required displacement sensitivity is set by the amplitude of the fibers' motion when the violin modes are highly excited.
For a given violin mode resonance, we can estimate the amplitude of the fiber's motion using the measured motion of the \ac{TM}. 
The TM acts as a simple harmonic oscillator with some amplitude $A_{TM}$. 
The fiber behaves as a standing wave with both ends clamped, having some antinode amplitude $A_{f}$. 
At the boundary where the TM and fiber attach, the transverse restoring force exerted by the tension in the fiber must match the inertial force of the TM at the resonant frequency, $\omega$.
This yields the following equivalence:

\begin{equation}
    A_{f}Tk\cos{(kL)}e^{i\omega t} = A_{TM}\omega^2 M e^{i\omega t}
\end{equation}

where $T$ is the tension in the fiber, $k$ is the wave number, $L$ is the length of the fiber, $t$ is time, and $M$ is the mass of the TM. 
The fiber is not perfectly clamped at its connection to the TM, so $kL$ is slightly off $n\pi$, but we take $|\cos{(kL)| \approx 1}$.
We also neglect the difference between the effective length and physical length of the fiber, as $M \gg m$.
Cancellation of terms and substitution of $\omega$ with the following dispersion relation:

\begin{equation}
    \omega  = k \sqrt{\frac{T}{\mu}} 
\end{equation}

where $\mu = \frac{m}{L}$, the mass per unit length of the fiber, and $k = \frac{n\pi}{L}$ for the $n$th violin mode (here $n=1$ for the fundamental), allows us to write:

\begin{equation}
    A_{f}  = \pi \frac{M}{m} A_{TM}
\end{equation}

where we see that multiplying the amplitude of the TM motion (which can be read straight from the calibrated \ac{DARM}) by the TM to fiber mass ratio allows us to estimate the amplitude of the fiber's motion. 
We now introduce the following parameters:

\begin{itemize}[noitemsep]
    \item Fiber diameter, $d_{fiber} = \SI{400}{\micro\metre}$
    \item Fiber length, $L_{fiber} = \SI{0.6}{\metre}$
    \item Density of fused silica, $\rho_{silica} = \SI{2200}{\kilo\gram\per\metre\cubed}$
    \item TM mass, $M = \SI{40}{\kilogram} $
    \item Amplitude of TM motion (VMs highly excited), $A_{TM} = \SI{1e-16}{\meter}_{RMS}$ \cite{vmodealog} 
\end{itemize}

with which we can estimate the amplitude of fiber motion to be \SI{7.5e-11}{m} RMS. 
Currently, the EVMD can sense a minimum RMS displacement of $\approx$ \SI{6e-12}{m}, better than the required sensitivity for highly excited violin modes. 
After being damped using current methods, typical excitation levels of the violin modes range between \SIrange{e-13}{e-15}{m}, defining the goal sensitivity regime for future \ac{EVMD} iterations.

\subsection{Damping requirement estimate}
\label{subsec:dampingest}

One can estimate the force required for damping by estimating the energy stored in the fiber, $E_{stored}$, and the damping quality factor, $Q_D$.
We assume a target energy damping time, $t_D$, of \SI{60}{\second} (the time for the stored energy to fall by a factor of $e$), at the violin mode frequency, $f_{f}$ of \SI{500}{\hertz}. 
In a highly excited state, the antinode amplitude of the fiber's motion, $A_{f}$, can reach up to \SI{7.5e-11}{\m} RMS or more.
Integrating over $\sin(kx)$ to account for the fiber's mode shape, the stored energy for an RMS antinode amplitude is:

\begin{eqnarray}
    E_{stored} = \frac{1}{2} m (\omega A_{f})^2 = \SI{4.6e-18}{\joule} \\
    Q_D = 2\pi t_D f_{fiber} = \num{1.9e5}
\end{eqnarray}

One can then determine the energy lost per cycle, $E_{lost}$:

\begin{equation}
    E_{lost} = 2\pi \frac{E_{stored}}{Q_D} = \SI{1.5e-22}{\joule}
\end{equation}

A force in phase with velocity does work $2\pi F_{rms} A_f$ per cycle, where $F$ and $A_f$ are both RMS values. 
Assuming the actuator is at the fiber antinode, equating this to $E_{lost}$ yields the required RMS force to damp the fiber:

\begin{eqnarray}
    F = \frac{E_{lost}}{2\pi A_{f}} \approx \boxed{\SI{3.2e-13}{\newton}}
\end{eqnarray}

The damping force requirements discussed above assume an energy damping time of \SI{60}{s}.
Realistically, modes can also be damped with a smaller force over a longer period of time.

\subsection{Electrostatic actuator damping force estimate}

The electrostatic actuator is modeled as a periodic plane exciter placed at $z = 0$, and the fused silica fiber is approximated as a dielectric bulk material at $z \ge h$. 
The exciter extends along the x-direction with periodicity $L$ and is infinite in the y-direction. 
The gap between adjacent electrodes has width $d$.
The potential of the electrodes is given by

\begin{equation}
    V(x) =
        \begin{cases}
            -\dfrac{V}{2} & -\dfrac{L-d}{2} \leq x \leq -\dfrac{d}{2}\\[6pt]
            \dfrac{V}{2} & \dfrac{d}{2} \leq x \leq \dfrac{L-d}{2}\\[6pt]
            \text{linear ramp between } \pm\dfrac{V}{2} & \text{elsewhere (fringing region)}
        \end{cases}
\end{equation}

With this potential, the Fourier components can be written as

\begin{equation}
    \label{eq:voltage_fourier}
V(x)= \sum_{n=0}^{\infty}V_n\sin{\frac{2\pi nx}{L}} ; \quad V_n = \frac{2V}{\pi n}\operatorname{sinc}{\phi} 
\end{equation}

with $\phi = \frac{d\pi n}{L} = \frac{dk_n}{2}$ and $k_n = \frac{2\pi n}{L}$.
We then apply Laplace's equation and a Fourier transform in the x direction:

\begin{equation}
\bigg(-k^2_n+\frac{\partial^2}{\partial z^2}\bigg)V_n(z)=0
\end{equation}

allowing us to write the electric field in the z direction at $z=h^+$ (inside the fiber) as 

\begin{equation}
    E_{n\perp}(h^+) = \frac{kV_n(z=0)}{\cosh(kh)+\epsilon_{fs}\sinh(kh)} 
\end{equation}

and the electric field in the z direction at $z=h^-$ (between the fiber and actuator) as 

\begin{equation}
    E_{n\perp}(h^-) = \frac{\epsilon_{fs}kV_n(z=0)}{\cosh(kh)+\epsilon_{fs}\sinh(kh)}
\end{equation}

where $\epsilon_{fs}$ is the relative permittivity of fused silica.
Assuming the fiber (approximated as a dielectric slab) has a charge area density of $\sigma$ at $z = h$ due to the surrounding field, we can calculate the force per unit area of the $n$-th Fourier component:

\begin{align}
    &f_n = \frac{\epsilon_0}{4}[E_{n\perp}^2(h^+)-E_{n\perp}^2(h^-)] \nonumber \\ 
    & = \frac{\epsilon_0}{4}(1-\epsilon_{fs}^2) \sum_{n=1,\mathrm{odd}}^{\infty} \bigg(\frac{k_n V_n(z=0)}{\cosh(k_n h)+\epsilon_{fs}\sinh(k_n h)}\bigg)^2
\end{align}

where $\epsilon_0$ is the vacuum permittivity. We can then plug in the Fourier coefficients and thus write

\begin{equation}
    \label{eq:force_from_model}
    F = V^2 \frac{4\epsilon_0(1-\epsilon^2_{fs})}{L^2}\sum_{n=1,\mathrm{odd}}^{\infty} \bigg( \frac{\mathrm{sinc}(\frac{\pi n d}{L})}{ \cosh(\frac{2\pi nh}{L})+\epsilon_{fs}\sinh(\frac{2\pi nh}{L}  )} \bigg)^2
\end{equation}

In the case of the electrostatic actuator, $V^2 = \Delta V^2 = |V_{s}-V_{b}|^2$, where $V_{b}\equiv V_{DC}$ is a large DC bias voltage that serves to induce polarization in the fiber. 
The polarized fiber can then be actuated upon by the signal voltage, $V_{s}$, which is an AC signal with amplitude $V_{AC}$ at the phase-shifted violin mode frequency, $\omega_{\text{VM}}$:

\begin{equation}
    V_{s} = V_{AC} \sin(\omega_{\text{VM}} t + \phi)
\end{equation}

Thus, the general force proportionality is

\begin{align}
    F = \alpha (\Delta V)^2 \nonumber\\
    &= \alpha\left(V_{DC}-V_{AC}\sin(\omega_{\text{VM}}t+\phi)\right)^2 \nonumber\\
    &= \alpha\underbrace{\left(V_{DC}^2 + \frac{V_{AC}^2}{2}\right)}_{\text{DC force}}
    \;-\;\alpha\underbrace{\frac{V_{AC}^2}{2}\cos(2\omega_{\text{VM}}t+2\phi)}_{\text{force at }2\omega_{\text{VM}}} \nonumber\\
    &\quad-\;\boxed{2\alpha\, V_{DC}V_{AC}\sin(\omega_{\text{VM}}t+\phi)}
\end{align}

where the boxed term is the AC force at the phase-shifted violin mode frequency, $\propto V_{DC}V_{AC}$, and $\alpha$ depends on the experimental and actuator design parameters $h$, $d$, $L$, and $\epsilon_{fs}$.

Because of the assumptions used to derive Eq. \ref{eq:voltage_fourier}, its accuracy for analytical modeling is limited within the constraints of the actuator's physical dimensions. 
However, this expression allows us to define the optimal electrode periodicity as $L=2\pi h$, thus informing design considerations for the actuator's geometry. 
In the initial prototype, the \ac{EVMD} actuator is constrained to a length of \SI{30}{\milli\meter} and width on the order of \SI{5}{\milli\meter}. 
Additionally, the previously mentioned buffer zone around the fiber must be maintained, setting $h=\SI{3}{\milli\meter}$. 

A preliminary actuator geometry, shown in Fig. \ref{fig:actuator}, was designed with the following parameters:

\begin{itemize}[noitemsep]
    \item Actuator length $= \SI{30}{\milli\meter}$
    \item Electrode periodicity, $L = \SI{5.5}{\milli\meter}$
    \item Electrode gap, $d = \SI{0.75}{\milli\meter}$
    \item DC Bias voltage, $V_{DC} = \SI{400}{\volt}$
    \item AC Signal voltage amplitude, $V_{AC} = \SI{10}{\volt}$
\end{itemize}

The initial periodicity of $L= \SI{5.5}{\milli\meter}$ was chosen as a middle ground between a low periodicity and the predicted optimum for preliminary experimental characterization. 
In future builds, the actuator length will increase to \SI{50}{\milli\meter}, and the exact optimal periodicity of $L=2\pi h$ will be implemented.

Modeling was performed analytically and using FEA to estimate the force this initial actuator can exert in the direction of the violin modes' oscillation.
Results from each are summarized in Table \ref{tab:force_comparison}.

\begin{table}[htbp]
    \vspace{6pt}
    \centering
    \begin{tabular}{l r}
        \toprule
        Modeling method & Estimated force (\si{\newton}) \\
        \midrule
        Analytical (using Eq.~\ref{eq:force_from_model})     & $-4 \times 10^{-10}$ \\
        Ansys Maxwell FEA      & $-3 \times 10^{-10}~$ \\
        \bottomrule
    \end{tabular}
    \caption{Comparison of RMS damping force estimates across modeling methods with increasing complexity, rounded to one significant figure. 
Discrepancies between different models are attributed to differing levels of complexity in each model, with the Ansys Maxwell FEA being closest to the physical system.
Negative forces correspond to attraction of the fiber.}
    \label{tab:force_comparison}
\end{table}
An analytical calculation using Eq.~\ref{eq:force_from_model} with the parameters above predicts an RMS force on the order of \SI{e-10}{\newton}.
Ansys Maxwell 3D CAD FEA was used for further assessment, modeling the exact actuator geometry shown in Fig. \ref{fig:actuator} alongside a silicon dioxide fiber (diameter \SI{400}{\micro\meter}, length \SI{0.6}{\meter}). 
Upon convergence, the model also predicts an RMS force on the order of \SI{e-10}{\newton}. 
Unlike expression \ref{eq:force_from_model}, Ansys captures the full 3D geometry of both the actuator and the fiber. 
The former method approximates the fiber as a dielectric slab, and assumes infinite electrode width in the direction transverse to the fiber length.

\vspace{2em}

\nocite{*}
\bibliography{bibliography}

\end{document}